\documentclass[12pt,epsf]{article}

\usepackage{epsfig,latexsym,amsfonts,amsmath,amsthm,amssymb,amsbsy,multirow,slashed,color,mathrsfs,subfigure,wrapfig}
\usepackage[font={footnotesize,sl},bf]{caption}

\usepackage{paralist}

\newcommand\be{\begin{equation}}
\newcommand\bea{\begin{eqnarray}}
\newcommand\ee{\end{equation}}
\newcommand\eea{\end{eqnarray}}

\newcommand{\bdm}{\begin{displaymath}}
\newcommand{\edm}{\end{displaymath}}

\newcommand{\f}[2]{\frac{#1}{#2}}
\newcommand{\bref}[1]{(\ref{#1})}

\usepackage{color}

\begin{document}
\begin{titlepage}
\hfill

\vspace*{20mm}
\begin{center}
{\Large \bf Limitations of holography}

\vspace*{15mm}
\vspace*{1mm}

Borun D. Chowdhury

\vspace*{1cm}

{Department of Physics, Arizona State University,\\
Tempe, AZ, 85287, USA \\
{\it bdchowdh@asu.edu}
}

\vspace*{1cm}

\end{center}

\begin{abstract}
By studying global $AdS$ using different foliations, global and Rindler-AdS, we show that there are two different asymptotic Fefferman-Graham expansions possible and thus two different definitions of ``boundaries''. We demonstrate that imposing boundary conditions on the two boundaries is not mutually compatible even when these boundaries are pushed to infinity. Thus, these two procedures define two genuinely distinct theories that we call global-CFT and Rindler-CFT. We show that the Rindler-CFT is not the same as the theory one gets by ``Rindlerizing the global-CFT'' described in~hep-th/9804085. We conjecture that the Rindler theory is incapable of capturing the dynamics inside the horizon and discuss its implications for the BTZ-CFT duality proposed in~hep-th/0106112.
\end{abstract}

\vskip 2cm

\begin{center}
Essay written for the Gravity Research Foundation \\ 2014 Awards for Essays on Gravitation.
\vskip 1cm

March 31, 2014
\end{center}
\end{titlepage}

\section*{Introduction}



One of the reasons AdS/CFT duality is so exciting is because of its potential to resolve the information paradox which stipulates that black holes violate unitarity. If the conjecture is correct, the unitarity of the CFT implies unitarity of the bulk. One hopes that the CFT will teach us what is wrong with Hawking's analysis that leads to the paradox~\cite{Hawking:1974sw}.
\begin{figure}[htbp] 
   \centering
   \includegraphics[width=3in]{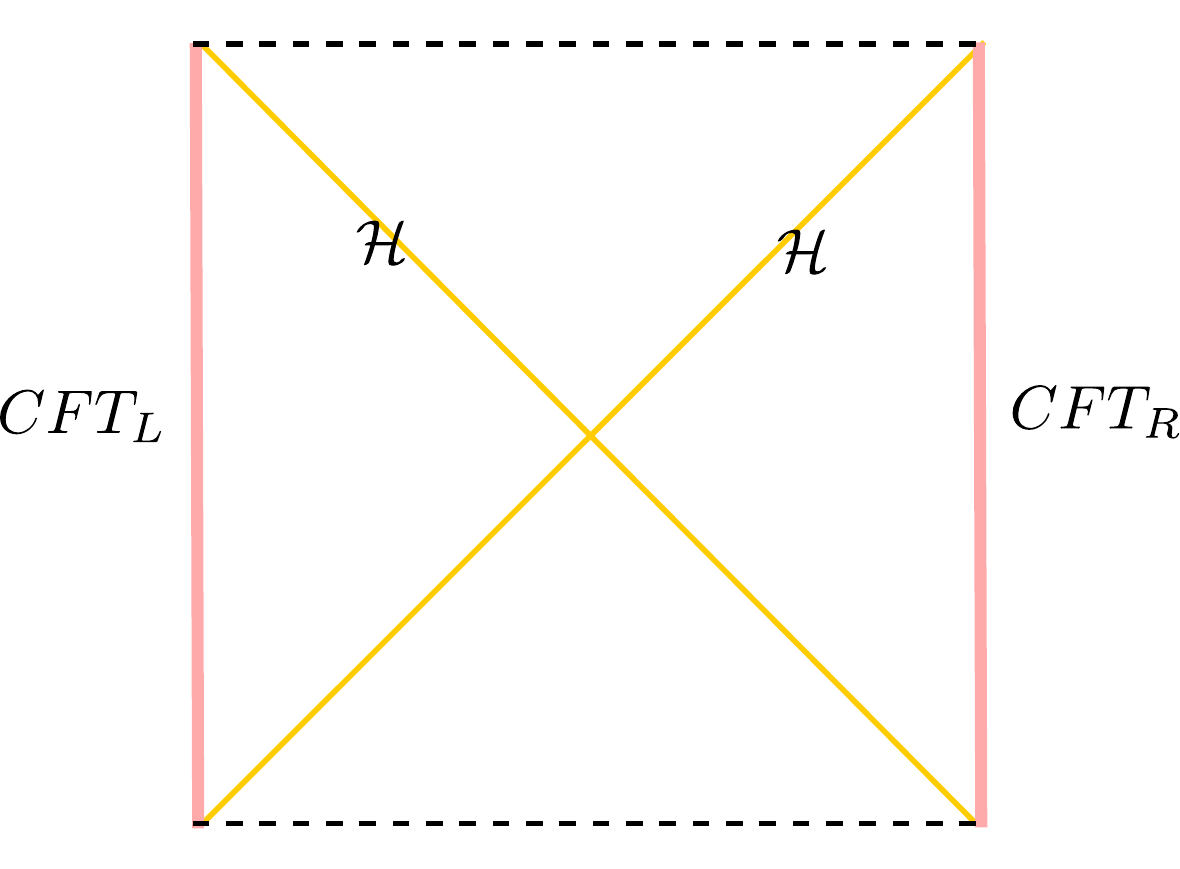}
    \caption{The eternal BTZ black hole has two asymptotically AdS boundaries. According to~\cite{Maldacena:2001kr} the bulk spacetime is dual to two CFTs living on these boundaries in a certain entangled state.}
   \label{fig:BTZ}
\end{figure}

If the CFT is to describe black holes, it must be able to ``see inside the horizon''. More rigorously, the dynamics behind the horizon should be captured by the CFT. A setting to investigate this is the purported duality between the BTZ black hole~\cite{Banados:1992wn} and two non-interacting two dimensional CFTs in a particular entangled state~\cite{Maldacena:2001kr} (see figure~\ref{fig:BTZ}). The CFTs are supposed to live on the two asymptotic boundaries of BTZ and may be compared to asymptotic observers outside a black hole who do not have access to the inside. Further, it has been argued that falling through the horizon requires an interaction between the two outside regions~\cite{Mathur:2012dx,Avery:2013bea,Mathur:2014dia} but the CFTs are non-interacting. This makes the BTZ-CFT duality suspect.

In this essay we will illustrate the limitations two decoupled CFTs have on the description of the bulk physics and discuss its implications for the proposed BTZ-CFT duality.

\section*{The dictionary between AdS and CFT}

To equate AdS and CFT one needs the dictionary between them. The first entry of this dictionary is the definition of the ``boundary" itself. Asymptotically AdS spacetimes permit the Fefferman-Graham expansion~\cite{FeffermanGraham} of their metrics
\be
ds^2 = dr^2 + \left(e^{2 r} g^{(0)}_{ab} + g^{(2)}_{ab} \right) dx^a dx^b + O(e^{-2r}).
\ee
The boundary is understood to be at a large fixed value of $r$.  This value $r_c$  is related to the cutoff of the dual theory and a CFT is obtained by taking $r_c \to \infty$.  $g^{(0)}_{ab}$ is the ``boundary metric" up to Weyl transformations. The on-shell variation of the gravity action (with a divergence cancelling counterterm) is
\be
\delta S \sim \int_{\partial M} d^2x  \sqrt{-g^{(0)}}~T^{ab} \delta g^{(0)}_{ab}
\ee
where $T^{ab}$ is a symmetric tensor that is interpreted as the CFT stress tensor~\cite{Balasubramanian:1999re}. The variational principle is well defined if we impose Dirichlet boundary conditions $\delta g^{(0)}_{ab}=0$.~\footnote{Some other boundary conditions are also allowed~\cite{Compere:2008us,Compere:2013bya,Avery:2013dja} but for brevity we only discuss Dirichlet boundary conditions.} Imposing a boundary condition specifies the theory and Dirichlet boundary conditions amount to holding the boundary fixed. $g^{(2)}_{ab}$ is allowed to fluctuate and carries information of the state.

\section*{Two Inequivalent CFTs}

\begin{figure}[htbp] 
   \centering
   \includegraphics[width=2in]{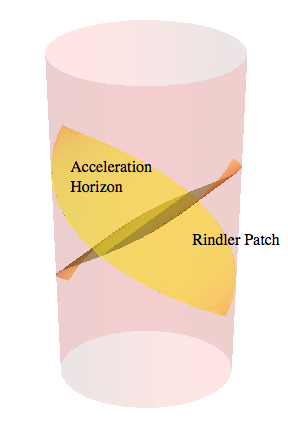}
    \caption{Global $AdS_3$ is depicted as a solid cylinder in red with $\tau$ going up. The Rindler coordinates cover the regions outside the acceleration horizons only. }
   \label{fig:GlobalAndRindler}
\end{figure}

One can write global $AdS_3$ metric as
\be
ds^2= d\rho^2 - \cosh^2 \rho~ d\tau^2 + \sinh^2 \rho~ d\phi^2 \label{globalAdS}
\ee
where $\rho \in [0,\infty]$, $\tau \in (-\infty,\infty)$ and $\phi \sim \phi + 2\pi$. These coordinates cover the entire manifold. It is useful to conformally compactify $\rho$ and visualise $AdS_3$  as a solid cylinder shown in figure~\ref{fig:GlobalAndRindler}.

 One can also write the metric in the so called AdS-Rindler coordinates
\be
ds^2 = d\eta^2 -\sinh^2 \eta~ dt^2 + \cosh^2 \eta~d\chi^2 \label{RindlerAdS}
\ee
where $\eta \in (0,\infty)$, $t \in (-\infty,\infty)$ and $\chi \in (-\infty,\infty)$. There is an acceleration horizon at $\eta=0$ and these coordinates cover the region outside the horizon. The rest of $AdS_3$ may be viewed as a Kruskal-like extension of these coordinates as shown in figure~\ref{fig:GlobalAndRindler}.

One can perform a large $\rho$ expansion to write  \bref{globalAdS} in the Fefferman-Graham form and define a field theory on the cylindrical boundary $S^1 \times R$ by holding the metric on the cutoff surface $\rho_c$ fixed. However, one can also perform a large $\eta$ expansion to write \bref{RindlerAdS} in the Fefferman-Graham form and define two field theories on $R^{1,1} \times R^{1,1}$ by holding the metric on the cutoff surfaces $\eta_c$  fixed. In either procedure, the cutoff surface has to be taken to infinity. We refer to the former CFT as the global-CFT and the latter CFT pair as the Rindler-CFTs.

It has been claimed in various places that the above two theories are the same. This is seen as some kind of CFT incarnation of  black hole complementarity(see~\cite{Barbon:2013nta} for example). We argue that this is not the case.

\begin{figure}[htbp] 
   \centering
   \subfigure[Finite cutoff surfaces]{
   \includegraphics[width=2in]{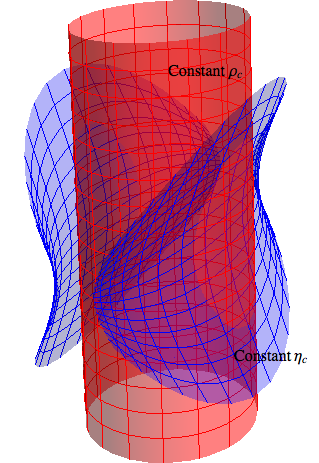}} 
   \hspace{1in}
   \subfigure[Infinite cutoff surfaces]{
   \includegraphics[width=2in]{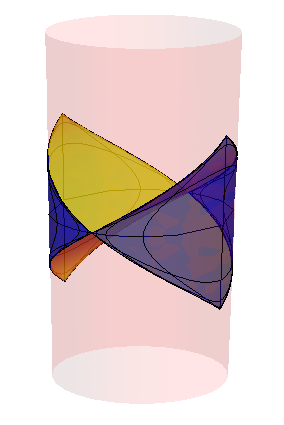}}
    \caption{(a) Global cutoff surface $\rho=\rho_c$ is shown in red and Rindler cutoff surface $\eta=\eta_c$ is shown in blue. Dirichlet boundary conditions on these are not compatible. (b) When we take $\rho_c$ to infinity, all the $\eta$ surfaces bunch up along the edges of the ``causal diamond". Two of them are shown in the figure.  Dirichlet boundary conditions on the cylinder amount to fixing the metric for all values of $\eta$ in a small interval near these edges.}
   \label{fig:CutoffSurfaces}
\end{figure}

In figure~\ref{fig:CutoffSurfaces}a we plot $\rho=\rho_c$  and $\eta=\eta_c$ surfaces. The two prescriptions described above require us to impose Dirichlet boundary conditions on these surfaces respectively. It is clear from this figure that these two are not consistent but one should take the cutoffs to infinity to be sure.

One can find the relation between the two coordinate systems in~\cite{Avery:2013bea} for example. We focus on one of those 
\be
\sinh^2 \rho= \sinh^2 \eta \left( \f{\cosh 2 \chi + \cosh 2 t}{2} \right)+ \sinh^2 \chi.
\ee

It is clear that for any given $\rho_c$, one gets  $\eta=0$  for large enough $\chi,t$. Said differently, the bulk acceleration horizon intersects  any cylinder of radius $\rho_c$ and imposing Dirichlet boundary conditions on the cylinder will always fix the metric for small $\eta$ when $\chi,t$ are large enough. This behaviour persists when  $\rho_c \to \infty$ and in this case the transition from large to small $\eta$ is squeezed into ever tinier intervals of $\delta \phi$ and  $\delta \tau$ but is always present nevertheless. This is shown in figure~\ref{fig:CutoffSurfaces}b.

If on the other hand one wants to define the Rindler-CFT then one needs to take large $\eta_c$ and permit arbitrarily large $\chi$ and $t$. Then one takes $\eta_c \to \infty$. This is clearly not consistent with the above procedure.

Thus, we have established that the Rindler-CFT obtained by taking $\eta_c$ large and fixing the corresponding $g^{\eta,(0)}_{ab}$ is different form the global-CFT obtained by taking $\rho_c$ large and fixing the corresponding $g^{\rho,(0)}_{ab}$  even with the cutoffs removed. 

\section*{Implications of inequivalence of Rindler and Global CFTs}

\begin{figure}[htbp] 
   \centering
   \subfigure[Rindlerized-global-CFT]{
   \includegraphics[width=2in]{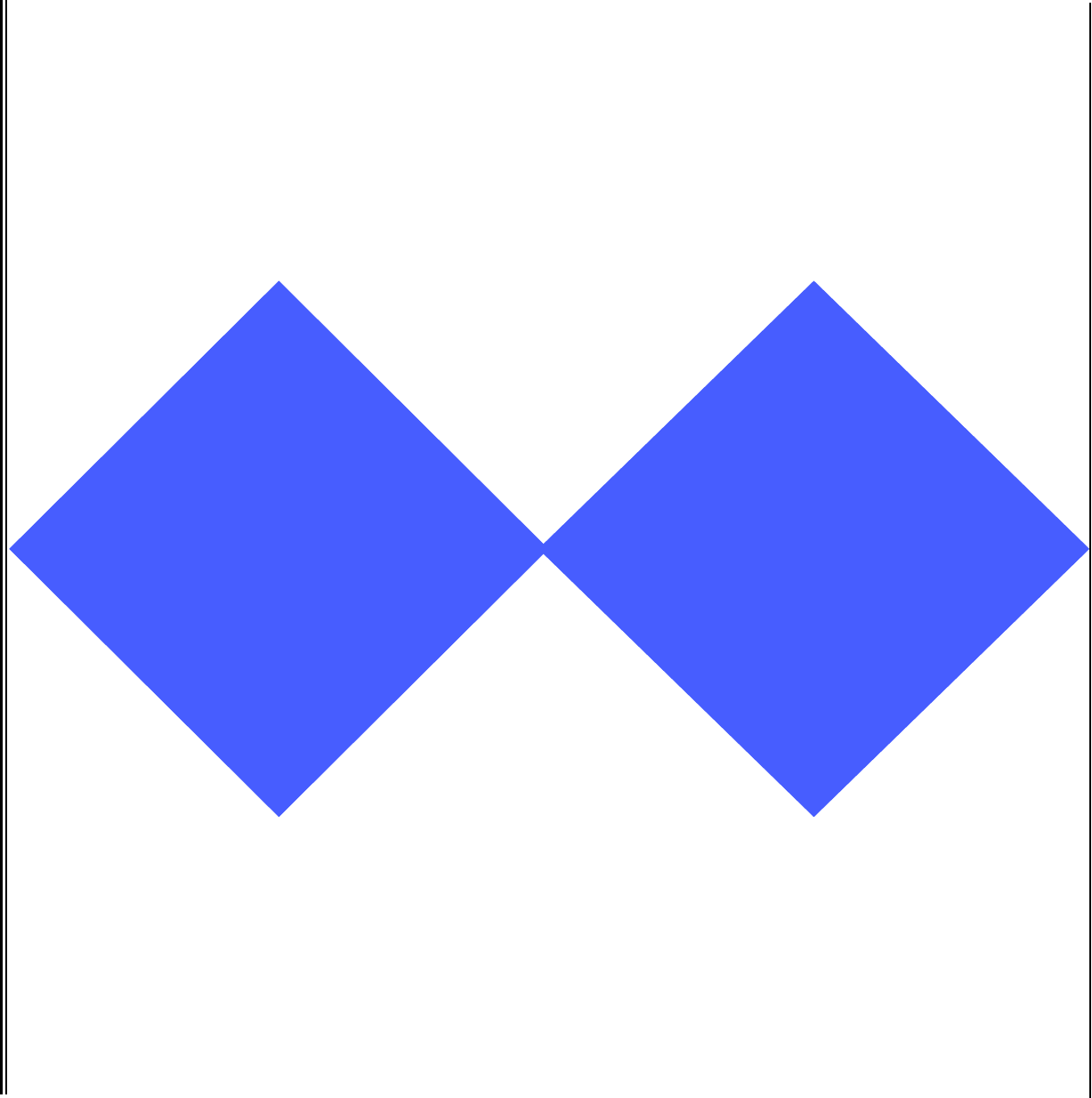}} \hspace{1in}
    \subfigure[Rindler-CFT]{
   \includegraphics[width=2in]{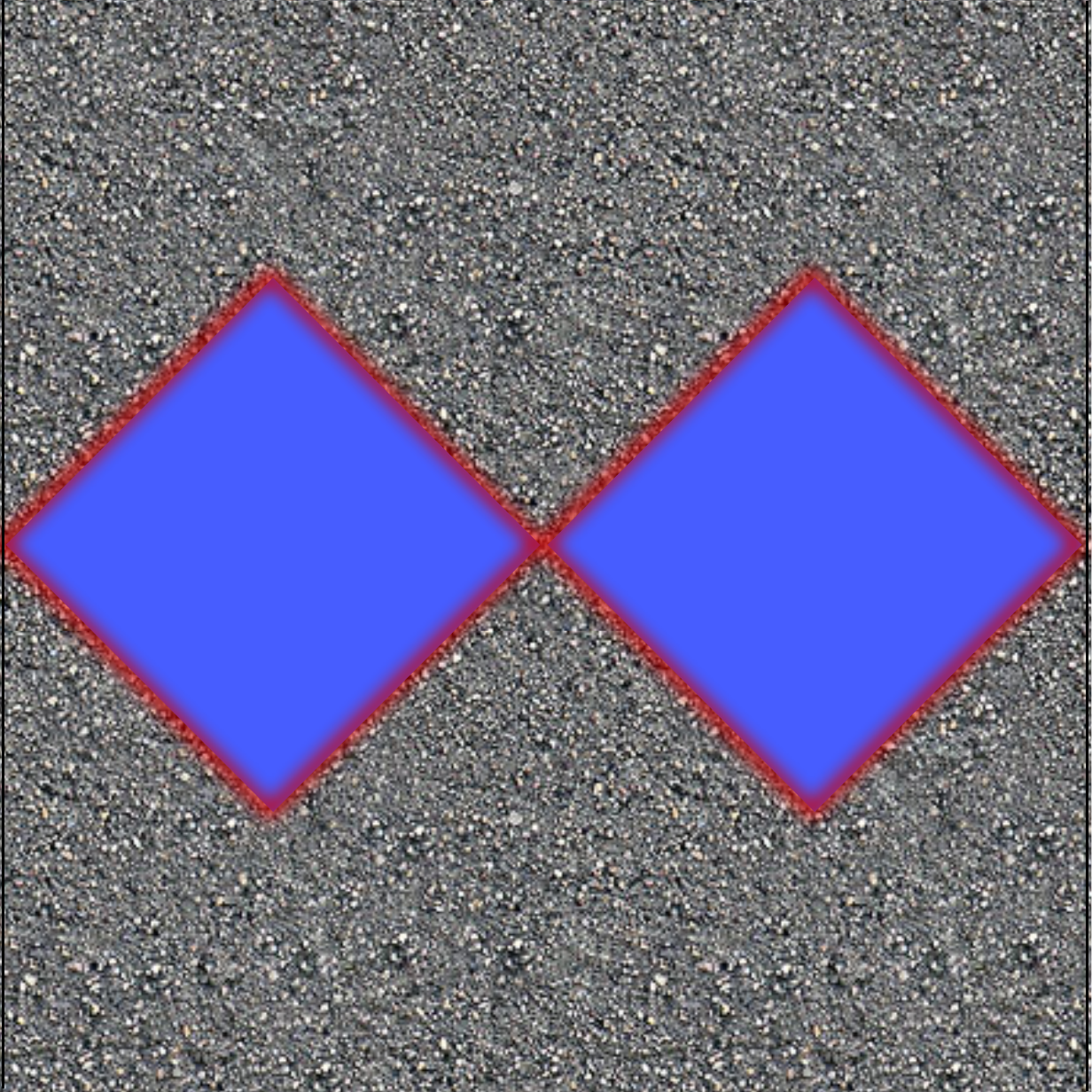}} 
    \caption{We open up the boundary cylinder for better visualisation. In (a) one formally divides the global-CFT into two halves and follows their causal development inside the ``causal diamonds''. This is analogous to ``Rindlerizing the global theory''~\cite{Maldacena:1998bw}. This is {\em not} the same as the Rindler-CFT (b). While the two theories are approximately equal deep inside the diamonds, they start differing at the edges. The global-CFT is defined everywhere but the Rindler-CFT is defined only  inside the diamonds.  }
   \label{fig:TwoCFTsInDiamonds}
\end{figure}

Though the global theory is defined on a cylinder, one may formally divide the cylinder into two halves at some  time and follow the causal development of each half (figure~\ref{fig:TwoCFTsInDiamonds}a) inside ``causal diamonds''. The coordinate transformation for this is obtained by  taking the large $\eta$ limit {\em first} and {\em then} projecting the resulting  transformations onto the global boundary. This formally defines two  CFTs  on $R^{1,1} \times R^{1,1}$ coming from ``Rindlerizing the global-CFT"~\cite{Maldacena:1998bw}. 

This set of CFTs is {\em not} the same as the Rindler-CFT discussed earlier. To see the difference between the two, note that the two halves of the global cylinder define two Hilbert spaces which have an interaction term between them in the Hamiltonian 
\be
H_{global} = H^{global}_L + H^{global}_R + H^{global}_{int} \label{GlobalHamiltonian}
\ee
which is invisible in Rindler evolution but is present nevertheless~\cite{Avery:2013bea}. On the other hand the Rindler-CFT is defined on two disconnected and decoupled manifolds $R^{1,1} \times R^{1,1}$ and thus define two Hilbert spaces which have no interaction term between them in the Hamiltonian 
\be
H_{Rindler} = H^{Rindler}_L + H^{Rindler}_R. \label{RindlerHamiltonian}
\ee

While the two CFTs, Rindlerized-global-CFT and Rindler-CFT, are approximately equal deep inside the causal diamonds in figure~\ref{fig:TwoCFTsInDiamonds}, they start differing near its edges. The global-CFT is defined outside the diamonds also by having Dirichlet boundary conditions but the Rindler-CFT is not defined outside the diamonds. In terms of correlation functions, it can be shown that the correlation functions of the two CFTs are approximately equal deep inside the diamond but start differing near the edges.

The implications of this difference are under investigation at the time of writing this essay. We nevertheless speculate a little on the same. Since the bulk acceleration horizons are not special, the global-CFT captures the dynamics both inside and outside these horizons. The Rindler-CFT, being different from the global-CFT, must imply different dynamics in the bulk. The edges of causal diamonds are projections of bulk horizons and based on the fact that the correlation functions of the two CFTs start differing there and that the Rindler CFT is not defined outside the diamonds, we conjecture that the Rindler-CFT is incapable of capturing the dynamics inside the horizon. In fact, based on~\cite{Mathur:2014dia,Kay:2013gia} it may be that the Rindler-CFT makes the bulk unstable and cuts it off outside the horizons.

\begin{figure}[htbp] 
   \centering
   \includegraphics[width=3in]{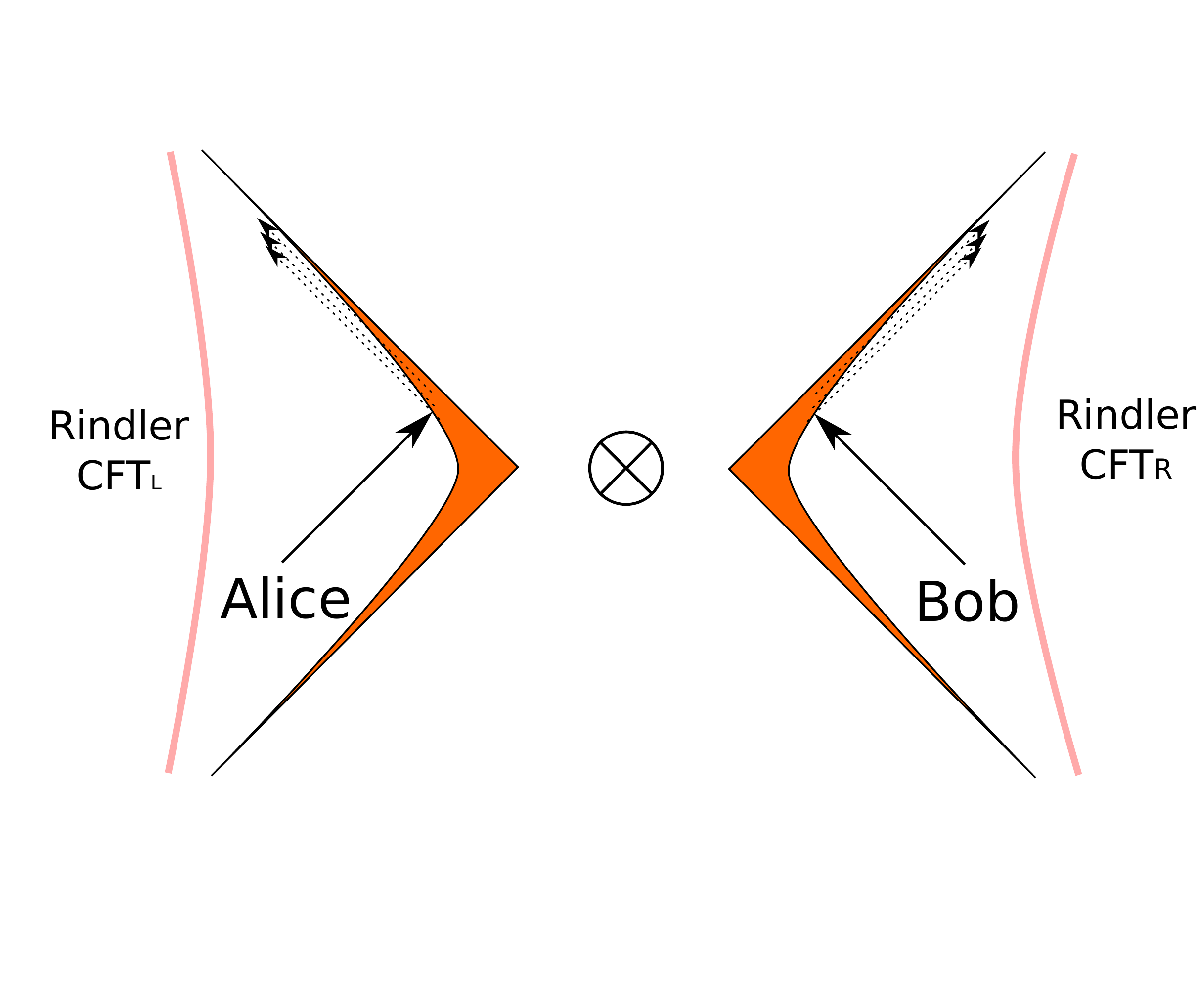}
    \caption{We conjecture that each exterior region is capped off outside the respective would-be horizons. Alice and Bob falling on the two sides thermalise on the fuzzy caps but do not meet each other. }
   \label{fig:OurPicture}
\end{figure}

As is well known the BTZ black hole is a quotient of $AdS_3$ obtained by  $\chi \sim \chi+2\pi$~\cite{Banados:1992gq}. This orbifolding is not a symmetry of the global or the Rindlerized-global theory but is a symmetry of the Rindler theory. Thus CFTs on the boundaries of BTZ must be obtained by orbifolding the Rindler-CFT and, if our conjecture is correct, can only describe physics outside the horizons and not inside and based on~\cite{Mathur:2014dia,Kay:2013gia} may imply the bulk ends in fuzzballs outside either horizon as shown in figure~\ref{fig:OurPicture}.

\section*{Acknowledgements}

I would like to thank Steve Avery,  Saugata Chatterjee,  Jan de Boer, Daniel Harlow, Andrew Long, Juan Maldacena, Samir Mathur, Lubos Motl, Kyriakos Papadodimas, Maulik Parikh, Joe Polchinski, Suvrat Raju, Eray Sabancilar, Masaki Shigemori, Tanmay Vachaspati and Erik Verlinde for valuable discussions.

\bibliographystyle{toine}
\bibliography{Final}

\end{document}